\documentclass
[prl,twocolumn,superscriptaddress,floatfix,10pt,nofootinbib,preprintnumbers]{revtex4}%
\usepackage{amsmath,amsfonts,setspace,cancel,url,hyperref,verbatim,slashed,amsthm,graphicx,color}
\newcommand{\be}{\begin{equation}}
\newcommand{\ee}{\end{equation}}

\newcommand{\gM}{\mathcal{M}}

\newcommand{\hmu}{{\hat{\mu}}}
\newcommand{\hnu}{{\hat{\nu}}}

\newcommand{\Gsix}{\mathrm{E}_6}

\newcommand{\ED}{E_{D(D)}} 

\newcommand{\gV}{{V}}

\begin{document}
\raggedbottom

\preprint{Imperial-TP-2017-ASA-02}

\title{Type II strings are Exceptional}

\author{Alex S. Arvanitakis}
\email{a.arvanitakis@imperial.ac.uk}
\affiliation{The Blackett Laboratory, Imperial College London, Prince Consort Road, London SW7 2AZ, U.K.}

\author{Chris D. A. Blair}
\email{cblair@vub.ac.be}
\affiliation{Theoretische Natuurkunde, Vrije Universiteit Brussel, and the International Solvay Institutes, Pleinlaan 2, B-1050 Brussels, Belgium}

\begin{abstract}

We construct the exceptional sigma model: a two-dimensional sigma model coupled to a supergravity background in a manifestly (formally) $\ED$-covariant manner.
This formulation of the background is provided by Exceptional Field Theory (EFT), which unites the metric and form fields of supergravity in $\ED$ multiplets before compactification.
The realisation of the symmetries of EFT on the worldsheet uniquely fixes the Weyl-invariant Lagrangian and allows us to relate our action to the usual type IIA fundamental string action and a form of the type IIB $(m,n)$ action. This uniqueness ``predicts'' the correct form of the couplings to gauge fields in both Neveu-Schwarz and Ramond sectors, without invoking supersymmetry.
\end{abstract}

\maketitle

The usual way to search for realistic lower-dimensional physics from string or M-theory is to compactify.
Early studies of the simplest reductions -- on tori -- led to the first encounter of T-duality \cite{Kikkawa:1984cp,Sakai:1985cs} in which string theory on a circle of radius $R$ is equivalent to string theory on a circle of radius $1/R$.
String theory on a $D$-torus leads to an $O(D,D; \mathbb{Z})$ T-duality symmetry, while M-theory on a $D$-torus has an $\ED(\mathbb{Z})$ U-duality involving the exceptional Lie groups. 
These are powerful tools for understanding how these theories are unified, and point to the idea that stringy or M-theoretic probes see spacetime geometry contrary to our usual expectations.


The search for reformulations of string theory in which the enlarged symmetries exhibited by these dualities are apparent before compactification 
begins with the construction of ``duality symmetric'' models \cite{Duff:1989tf, Tseytlin:1990nb, Tseytlin:1990va, Hull:2004in, Hull:2006va}.
Here, the sigma model's target space is \emph{doubled} and the coordinates appearing on the worldsheet are the doubled pair $(Y, \tilde Y)$. T-duality then swaps $\tilde Y$ for $Y$. A chirality constraint relates the $\tilde Y$ back to the $Y$, so that the number of degrees of freedom is not increased. 
Similar ideas were pioneered for membranes in \cite{Duff:1990hn}, but this approach runs into some difficulties \cite{Duff:2015jka}. 

We can also pursue this problem in the supergravity picture.
Here, Double Field Theory (DFT) \cite{Siegel:1993th, Siegel:1993xq, Hull:2009mi, Hull:2009zb, Hohm:2010jy, Hohm:2010pp}
and Exceptional Field Theory (EFT) 
\cite{Berman:2010is, Berman:2011jh,Berman:2011cg, Berman:2012vc, Hohm:2013vpa, Hohm:2013uia, Hohm:2014fxa, Godazgar:2014nqa, Hohm:2015xna, Abzalov:2015ega, Musaev:2015ces, Berman:2015rcc}
reformulate 10- or 11-dimensional supergravity with all bosonic fields in representations of $O(D,D)$ or $\ED$ (the fermions appear in representations of the maximally compact subgroups), depending on an extended set of coordinates $(X^\mu, Y^M)$. For DFT, the $Y^M$ are simply doubled coordinates, while for EFT, more extra coordinates are needed such that the $Y^M$ fit into a particular representation, denoted $R_1$, of $\ED$ (so for $O(D,D)$ $R_1$ is the fundamental). 

The extended space parametrised by the $Y^M$ comes with a \emph{local} $O(D,D)$ or $\ED$ symmetry provided by ``generalised diffeomorphisms'' (which combine ordinary diffeomorphisms and gauge transformations).
These are generated by generalised vectors $\Lambda^M$, which act on another generalised vector $U^M$ via the generalised Lie derivative: $ \delta_\Lambda U^M = \mathcal{L}_\Lambda U^M$ with
\be
\mathcal{L}_\Lambda U^M \\
 = \Lambda^N \partial_N U^M - U^N \partial_N \Lambda^M + Y^{MN}_{PQ} \partial_N \Lambda^P U^Q \,,
 \label{gendiff}
\ee
where the deviation from the usual form of the Lie derivative is due to the final term involving $Y^{MN}_{PQ}$ which is constructed from $O(D,D)$ or $\ED$ invariant tensors (see \cite{Berman:2012vc}).
Now, coordinate dependence in principle can be on any of the $Y^M$, but 
this is in fact constrained, as follows from closure of algebra of generalised diffeomorphisms, for which we impose 
the ``section condition''
\be
\label{sectioncondition}
Y^{MN}_{PQ}\partial_M \otimes \partial_N=0\,.
\ee
A choice of physical $Y^i$ coordinates satisfying this is a choice of ``section''. Such a choice breaks $O(D,D)$ or $\ED$ and establishes the link to the usual formulation of supergravity without extended coordinates. When isometries are present, there is an ambiguity in the choice of section. This corresponds to the usual notion of duality.

We can think of the  ``doubled'' sigma model of \cite{Hull:2004in, Hull:2006va} as describing strings propagating on a DFT background; upon eliminating dual coordinates the DFT background reduces to a standard supergravity one and the doubled sigma model reduces to the conventional sigma model.

We will provide a similar worldsheet picture for $\ED$ ($D\leq 6$), which we call the exceptional sigma model. 
This describes a string coupling to an EFT background.
This background will consist of the following EFT tensors.
Firstly, there are metric-like degrees of freedom: the ``external'' metric, $g_{\mu\nu}$ (roughly, the metric involving only the $X^\mu$ coordinates, which do not transform under $\ED$), and the ``generalised metric'' $\gM_{MN}$ (roughly, the metric involving only the $Y^M$ coordinates). 
Secondly, we have generalised gauge fields, including a one-form $A_\mu^M$ in the $R_1$ representation of $\ED$, and a two-form $B_{\mu\nu}$, in another representation of $\ED$ denoted by $R_2$. 
These are the first two fields in the ``tensor hierarchy'' of EFT \cite{Cederwall:2013naa, Wang:2015hca}.
The representation $R_2$ is always contained within the symmetric product of two $R_1$ representations, and so we can write the field $B_{\mu\nu} \in R_2$ as carrying a pair of symmetric $R_1$ indices, thus $B_{\mu\nu}^{MN}$ (symmetrisation implicit). 

Unsurprisingly, the representations $R_1, R_2,\dots$ that characterise these form fields are exactly the representations into which the brane ensemble of string/M-theory reassembles upon toroidal reduction (see e.g. \cite{Obers:1998fb}); for instance, upon reducing on a $T^D$  
M2 and M5 branes completely wrapping the torus directions appear as particles in the reduced theory, transforming in the $R_1$ representation. M2 and M5 branes with one worldvolume direction unwrapped appear as strings -- transforming in the $R_2$ multiplet. And so on. The conceptual difficulty is that different kinds of branes are mapped to each other by the action of $\ED$. 
A way around this is to construct $(p-1)$-brane actions coupling to the $p$-form in $R_p$, which in 10- or 11-dimensions describe only the genuine $(p-1)$-branes that occur there, but which can be interpreted in lower dimensions as describing the full multiplet of wrapped branes. This is the logic of the EFT particle actions ($p=1$) studied in \cite{Blair:2017gwn}. (Alternative approaches to U-duality covariant branes include \cite{Bengtsson:2004nj,Linch:2015fya,Linch:2015qva, Linch:2015fca, Sakatani:2016sko}). 

We now present the action. To couple the multiplet $B_{\mu\nu}^{MN}$ we introduce a set of charges $q_{MN}$ (valued in the representation $\bar R_2$ of $\ED$ inside the symmetric tensor product $\bar R_1\otimes \bar R_1$). We denote the worldsheet coordinates by $\sigma^\alpha$, the worldsheet metric by $\gamma_{\alpha \beta}$ and the Levi-Civita symbol by $\epsilon^{\alpha \beta}$.
The extended spacetime coordinates appear as worldsheet scalars $(X^\mu(\sigma) , Y^M(\sigma))$, and the background fields can depend on these subject to the section condition. 
We also need an auxiliary worldsheet one-form $V_\alpha^M$ which appears in the covariant worldsheet differential
\be
\label{DY}
D_\alpha Y^M d\sigma^\alpha = ( \partial_\alpha Y^M +  A_\alpha^M + V_\alpha^M) d\sigma^\alpha \,,
\ee
in which the EFT 1-form $A_\mu^M$ also appears (we write $A_\alpha^M \equiv \partial_\alpha X^\mu A_\mu^M$). This field $V_\alpha^M$ essentially serves to gauge away the dual coordinates. 
Consider splitting $Y^M = ( Y^i , Y^A)$, such that $Y^i$ are physical and the $Y^A$ are dual, one obtains a shift symmetry in the $Y^A$ (as the section condition is solved by $\partial_i \neq 0$, $\partial_A = 0$).
In \cite{Hull:2004in, Hull:2006va}, gauging this symmetry allows one to eliminate the $Y^A$ from the action. 
For this to work, we require
\be
\gV_\alpha^M \partial_M = 0 \,,
\ee
so that for the section $\partial_i \neq 0$, $\partial_A = 0$, only the components $V^A$ are present. 

The action is then given by $S = - \frac{1}{2} \int d^2\sigma ( L_{\rm kin} + L_{\rm WZ} )$ with
\begin{align}
L_{\rm kin}&=  
T\sqrt{-\gamma}\gamma^{\alpha \beta}\label{kin}\\
\times&\Big(\frac{1}{2}\mathcal M_{MN} D_\alpha Y^M D_\beta Y^N
  +g_{\mu\nu}\partial_\alpha X^\mu \partial_\beta X^\nu\Big) \,,
\nonumber\\
L_{\rm WZ} &=  q_{MN} \epsilon^{\alpha \beta} \left(
 B_{\alpha\beta} ^{MN} + A_\alpha^M D_\beta Y^N+\partial_\alpha Y^M V_\beta^N 
\right)  
\label{WZ}
\end{align}
where $B_{\alpha\beta}^{MN}=B_{\mu\nu}^{MN}\partial_\alpha X^\mu \partial_\beta X^\nu$
and $T$ (the ``tension'') is
\be
\label{tension}
T = \sqrt{\frac{1}{2(D-1)} \gM^{MN} \gM^{PQ} q_{MP} q_{NQ}} \,
\ee
As we will explain, the construction of this action, coupling in a natural -- and in particular gauge-invariant -- manner to the two-form $B_{\mu\nu}$ of the EFT tensor hierarchy, is exceptionally constrained by the requirement of invariance under the intricate $\ED$ local symmetries of EFT, and leads us to the \emph{unique result} \eqref{kin}, \eqref{WZ}, \eqref{tension} (modulo some reasonable assumptions). 
Gauge invariance also restricts the choice of $q_{MN}$: for a generic 10-dimensional background (i.e. a background dependent on $D-1$ of the $Y^M$), one finds that $q_{MN}$ can only select the strings known to exist in ten dimensions. 
Remarkably, this includes the correct couplings to the 10-dimensional two-forms, otherwise fixed by supersymmetry.

The worldsheet one-form $V_\alpha^M$ also ensures covariance under the local symmetries of EFT. As pointed out in \cite{Park:2013mpa, Lee:2013hma}, the natural candidate kinetic term $\gM_{MN} D_\alpha Y^M D_\beta Y^N$ involving the generalised metric on the extended space only transforms properly when $\gV_\alpha^M$ is present and assigned a particular transformation under the local symmetries of DFT. This generalises naturally to EFT \cite{Blair:2017gwn}. 
Consider first the $\ED$ generalised diffeomorphisms, defined in \eqref{gendiff}.
These act on the generalised metric as a tensor, and as gauge transformations of $A_\mu^M$ via $\delta_\Lambda A_\mu^M = \partial_\mu \Lambda^M - \mathcal{L}_{A_\mu} \Lambda$.
The worldsheet action should obey a covariance requirement under generalised diffeomorphisms: namely, varying the coordinates on the worldsheet as $\bar \delta_\Lambda Y^M = \Lambda^M(X,Y)$ should \emph{induce} the correct transformations $\delta_\Lambda$ of the background fields. This is then a symmetry of the worldsheet only if $\Lambda^M$ is a generalised Killing vector, i.e. $\delta_\Lambda = 0$ on all background fields. 
In addition, $A_\mu^M$ also transforms under one-form gauge transformations valued in $R_2$, as $\delta_\lambda A_\mu^M = - Y^{MN}_{PQ} \partial_N \lambda_\mu^{PQ}$, and the worldsheet action should be invariant under such gauge transformations.
Imposing these requirements on the generalised metric coupling $\gM_{MN} D_\alpha Y^M D_\beta Y^N$ fixes the transformations of the gauge field $V_\alpha^M$ to be
\be
\begin{split} 
\bar \delta_\Lambda V_\alpha^M & = - Y^{MN}_{PQ} ( \partial_N \Lambda^P D_\alpha Y^Q + \partial_N A_\mu^P \partial_\alpha X^\mu \Lambda^Q ) \,,\\
\delta_\lambda V_\alpha^M & = Y^{MN}_{PQ} \partial_N \lambda_\mu^{PQ} \partial_\alpha X^\mu \,.\\
\end{split} 
\ee
Note that these preserve $V_\alpha^M \partial_M = 0$ (using the section condition \eqref{sectioncondition}). Actually, the presence of certain weight terms in the generalised Lie derivative acting on the generalised metric in fact force us to introduce $T$ as defined in \eqref{tension} such that altogether it is the combination $T\gM_{MN} D_\alpha Y^M D_\beta Y^N$ which obeys the covariance requirement.

One can use this information to then construct the gauge invariant completion of the electric WZ coupling, beginning with the gauge transformation of $B_{\mu\nu}$, which is
\be
\delta_\lambda B_{\mu\nu}^{MN} = 2 \partial_{[\mu} \lambda_{\nu]}^{MN}- \mathcal{L}_{A_{[\mu}} \lambda_{\nu]}^{MN} +\frac{1}{2(D-1)} Y^{MN}_{PQ} A_{[\mu}^P \delta_\lambda A_{\nu]}^Q\,,
\ee
with the end result being \eqref{WZ}.

As mentioned before, the WZ coupling is only gauge invariant if $q_{MN}$ is constrained:
\be
q_{MN} Y^{NP}_{KL} \partial_P = q_{KL} \partial_M  \,.
\label{magic}
\ee
(this arises from considering $V$ and $A$ independent terms in the gauge transformation of \eqref{WZ}.)
The idea is to solve this constraint for the charge $q_{MN}$ after imposing the section condition $\partial_i \neq 0$, $\partial_A = 0$.
(The role of this constrained charge in simultaneously ensuring gauge invariance and selecting the allowed branes appears to be a generic feature of brane formulations in EFT, as has been proposed in \cite{CederwallKorea, MalekTalk, BCM}.)
Generically, there are no solutions for the section choice which relates EFT to 11-dimensional supergravity -- unless one of the physical directions $Y^i$ is an isometry. This reduces us to 10-dimensional type IIA, and we find that there is a single solution corresponding to the single F1 string of type IIA.
On the type IIB sections, one finds intead that there is a doublet of allowed solutions transforming under the unbroken $\mathrm{SL}(2) \subset \ED$: this corresponds to the $(m,n)$ strings of type IIB. 

So far we have constructed the WZ coupling and the generalised metric pullback $T \gM_{MN}D_\alpha Y^M D_\beta Y^N$ by imposing gauge invariance under the EFT $B$-field gauge transformations (with parameter $\lambda^{MN}_\mu$) and generalised diffeomorphisms (with parameter $\Lambda^M$). We can also write down the pullback of the external metric $T g_{\mu\nu}\partial_\alpha X^\mu \partial_\beta X^\nu$ which automatically respects both symmetries. It remains to consider the EFT ``external'' diffeomorphisms with parameter $\xi^\mu$.

Remarkably, imposing external diffeomorphism covariance requires an interplay between the kinetic and WZ pieces, thereby fixing all but one relative coefficient. (Again the covariance requirement is that under a coordinate transformation on the worldsheet, $\bar\delta_\xi X^\mu = \xi^\mu(X,Y)$, one should find induced the usual spacetime transformations $\delta_\xi$ of the background fields, which is then a global symmetry if $\xi^\mu$ is a Killing vector. One needs as before to include a transformation of $V_\alpha^M$ in order that this works.) This interplay follows inescapably from the following piece in the transformation of $A_\mu^M$:
\be
\delta_\xi A_\mu{}^M \supset \gM^{MN} g_{\mu\nu} \partial_N \xi^\nu
\ee
which must appear on varying the WZ term. However, there is no way of generating this as no other $\gM$-dependent terms appear in the variation of $L_{\rm WZ}$.
This suggests we must be able to obtain it from the kinetic term. In practical terms, the calculation leads to the following anomalous variation, which must vanish:
\begin{align}
 -\frac{1}{2}\int d^2\sigma \, &g_{\mu\nu} \partial_K \xi^\mu \partial_\alpha X^\nu \gM^{KM} \label{deltaxiS}
\\ &\times ( T \sqrt{-\gamma} \gamma^{\alpha \beta} \gM_{MN} D_\beta Y^K - q_{MN} \epsilon^{\alpha \beta} D_\beta Y^N ) \,,\nonumber
\end{align}
This can be compared with the variation of the action with respect to $V_\alpha^M$:
\be
\begin{split} 
\delta S  = -\frac{1}{2}& \int d^2 \sigma \,\delta V_\alpha^M \big[ T \sqrt{-\gamma} \gamma^{\alpha \beta}\gM_{MN} D_\beta Y^N 
 \\ & \qquad\qquad - \epsilon^{\alpha \beta} q_{MN} ( D_\beta Y^N - V_\beta^N) \big]\,.
\end{split} 
\label{deltaVS}
\ee
Solving the section condition so that $\partial_i \neq0$, $\partial_A = 0$, we know that only $V_\alpha^A$ appears. 
It turns out that the only non-zero components of the charge allowed by \eqref{magic} are $q_{Ai}= q_{iA}$, using this and the equations of motion for the $V_\alpha^A$ components, one can show group-by-group that \eqref{deltaxiS} vanishes upon inserting the standard parametrisations for the generalised metric on section. 
Alternatively, one can cancel \eqref{deltaxiS} off-shell by including a further transformation of $V_\alpha^M$:
\be\bar
\delta_\xi \gV_\alpha^M \supset- \frac{1}{T \sqrt{-\gamma}} \gamma_{\alpha \beta} \epsilon^{\beta \gamma} \gM^{MP} q_{PQ} \gM^{QK} \partial_K \xi^\mu g_{\mu\nu} \partial_\gamma X^\nu  
\ee
For this to work, some miraculous identities must hold involving the charge $q_{MN}$: we need
\be
\gM^{MP} q_{PQ} \gM^{QN} \partial_M \otimes \partial_N = 0 \,,
\label{id1}
\ee
\be
\Big( \delta^K_Q - \frac{1}{T^2} \gM^{KM} q_{MN} \gM^{NP} q_{PQ} \Big) \partial_K = 0 
\label{id2}
\ee
(where the latter can also be viewed as fixing the precise form of $T$).
These follow from the magic requirement \eqref{magic}, as can be checked on a group by group basis. 

The reader familiar with the doubled sigma model may find these identities appear somewhat familiar. Let us explain.
In fact, our action, \eqref{kin} and \eqref{WZ}, also describes the doubled sigma model in the formulation of \cite{Hull:2004in, Hull:2006va, Lee:2013hma} (the $D-1$ in the tension \eqref{tension} corresponds to $O(D-1,D-1)$). In this case one has $q_{MN} = T_{F1} \eta_{MN}$, where $T_{F1}$ is the tension of the fundamental string, and $\eta_{MN}$ is the $O(D-1,D-1)$ structure. One has $Y^{MN}_{PQ} = \eta^{MN} \eta_{PQ}$ so the requirement \eqref{magic} is identically satisfied -- implying there is always a doubled string. The identities \eqref{id1} and \eqref{id2} are just the statement that $S^2 = I$ for $S^M_N = \eta^{MN} \gM_{NP}$ (saying that the generalised metric is an element of $O(D-1,D-1)$). In that case, they hold without contractions with derivatives and imply the consistency of the ``twisted self-duality'' constraint $DY^M = \star S^M_N DY^N$ which in turn kills the anomalous variation \eqref{deltaxiS} when $V_\alpha^M$ is on-shell. In EFT, the identities only hold upon contractions with derivatives, but a directly analogous twisted self-duality constraint is true --- that is, if we fix the last remaining relative coefficient in the exceptional sigma model lagrangian. This is explained in \cite{Arvanitakis:2018hfn}.

We have just described the construction of the exceptional sigma model based on the local symmetries of EFT. It is curious to note how similar the procedure is to the usual method for obtaining the spacetime EFT action \cite{Hohm:2013pua}, where a collection of terms which are separately invariant under generalised diffeomorphisms have their relative coefficients determined by requiring invariance under the external diffeomorphisms, with intricate interplay of ``gauge''- and ``metric''-like terms. The details 
will be reported 
in \cite{Arvanitakis:2018hfn}, including verifying uniqueness of the resulting action. We will also discuss the inclusion of a Fradkin-Tseytlin term in which $T$ as in \eqref{tension} plays the role of a generalised dilaton. 

We now outline how our action reproduces the type II string and D1-brane actions. The procedure is similar to that for the doubled sigma model \cite{Hull:2004in, Hull:2006va, Lee:2013hma}. 
On choosing a solution for the section condition as before, we impose the algebraic equation of motion for the non-zero components $V_\alpha^A$. This determines $D_\alpha Y^A$ in terms of $D_\alpha Y^i$. 
In order to do so, we should solve \eqref{magic} for the allowed non-zero components of the charge $q_{MN}$, finding in general that $q_{AB} = q_{ij} = 0$.
Then, using the dictionary relating the EFT fields to components of the 10-dimensional supergravity fields we find the action reduces to that of the IIA F1 or the IIB $(m,n)$ string, up to a single term involving the dual coordinates $Y^A$: 
\be
 - \frac{1}{2} \int d^2\sigma \epsilon^{\alpha \beta} q_{Ai} \partial_\alpha Y^A \partial_\beta Y^i \,,
\ee
which is a total derivative. Something similar appears in the doubled sigma model, and is cancelled by adding a so-called topological term, which in fact ensures the quantum consistency of the model \cite{Hull:2004in, Hull:2006va, Berman:2007vi}. For $O(D,D)$, this involves an antisymmetric tensor $\Omega_{MN}$, which can be interpreted as a symplectic term on the doubled space. Ordinarily this is not included in DFT, but it plays a central role in the related proposals of \cite{Freidel:2013zga, Freidel:2015pka, Freidel:2017yuv}.

Let us briefly discuss the $E_{6(6)}$ EFT, for which dictionaries relating the EFT fields to supergravity ones are provided in \cite{Hohm:2013vpa, Baguet:2015xha}. 
The representation $R_1$ is the 27-dimensional fundamental, and the $R_2$ representation is its conjugate.
There are two totally symmetric invariant tensors: $d^{MNP}$ and $d_{MNP}$.
A field $B_M \in R_2$ can be written with two upper indices as $B^{MN} = d^{MNP} B_P$, leading to the identification $q_{MN} = d_{MNP} q^P$.
The Y-tensor is $Y^{MN}_{PQ} = 10 d^{MNK} d_{PQ K}$. In a IIB section we split $E_{6(6)} \rightarrow \mathrm{SL}(5) \times \mathrm{SL}(2)$, in which $Y^M = ( Y^i, Y_{i a} , Y^{[ij]} , Y_a)$, with $i,j=1,\dots,5$ and $a,b=1,2$. One then finds that the condition \eqref{magic} kills all components of $q^M$ except the ${\rm SL}(2)$ doublet $q_a$. After some work we find
the Lagrangian becomes that of a IIB $(m,n)$ string:
\be
T_{F1} 
\tau_{m,n} \sqrt{-\gamma}\gamma^{\alpha \beta} \hat g_{\hmu \hnu}  \partial_\alpha X^{\hmu} \partial_\beta X^{\hnu} 
 +
 \epsilon^{\alpha\beta} q_a \hat C_{\hmu \hnu}{}^a \partial_\alpha X^{\hmu} \partial_\beta X^{\hnu}
\ee
where the ${\rm SL}(2)$ doublet $q_a$ is straightforwardly related to $(m,n)$ through $q_a=\sqrt{10} T_{F1} (m,n)$, $\tau_{m,n} = \sqrt{  e^{- 2 \Phi}  n^2 + ( m+ C_{(0)} n )^2}$, $\hat g_{\hmu \hnu}$ is the 10-dimensional string frame metric and $\hat C_{\hmu \hnu}{}^a$ is the doublet of 10-dimensional RR and NSNS 2-forms. The 10-dimensional coordinates are $X^{\hmu} = ( X^\mu, Y^i)$. 
For $(m,n) = (1,0)$, this is the F1 action, for $(m,n) = (0,1)$ we see the tension scales with $g_s^{-1}$ as expected for the D1. 
The action for general $(m,n)$ is related to the F1 action by an S-duality transformation and to the usual D1 action by integrating out the BI vector \cite{Schmidhuber:1996fy}.
It can also be obtained from the $\mathrm{SL}(2)$ covariant formulation of \cite{Townsend:1997kr, Cederwall:1997ts}, which can be viewed as a precursor to our exceptional sigma model.

Similarly, one can obtain the IIA fundamental string by working with a IIA solution of the section condition. If viewed as a reduction of an M-theory section, for which $Y^M = ( Y^i ,Y_{[ij]}, Y^{\bar i})$ where $i$ and $\bar i$ are 6-dimensional indices and $\partial_i \neq 0$, then one obtains the IIA section whenever there is no dependence on one of the M-theory coordinates, say $\partial_1 \neq 0$. In this case, the single non-zero charge allowed by \eqref{magic} is $q^{\bar 1} \propto T_{F1}$. 

Therefore the unique exceptional sigma model action can be reduced to an action for the standard 1-branes of type IIA and type IIB string theory, on solving the section condition for these cases and eliminating the gauge field $V_\alpha^M$. 
This requires the constraint \eqref{magic} on the charges $q_{MN}$ appearing in the WZ coupling of the two-form. For the 10-dimensional IIA and IIB sections, the number of solutions of this constraint is 1 and 2 respectively, leading inevitably to the IIA fundamental string and IIB F1/D1 bound state \cite{Witten:1995im}. 

One could also reduce the action below 10-dimensions. The obstacles to gauge invariance/covariance all vanish when $\partial_M = 0$, in which case $q_{MN}$ is unconstrained by \eqref{magic}.
Here \emph{the natural conjecture is that the exceptional sigma model describes the $\ED$ string multiplet in $11-D$ dimensions obtained by toroidal reduction.} For instance, the ${\rm SL}(5)$ exceptional sigma model lagrangian on a background with $\partial_M=0$ should describe the string quintuplet in seven dimensions consisting of the four M2 branes wrapped around a single compactified dimension along with the M5 wrapping all four compactified dimensions.

The methods we used to construct the action can be systematically applied to study branes in EFT.
One application of the doubled sigma model is to define and study strings in T-fold backgrounds \cite{Hull:2004in}, suggesting the exceptional sigma model would be relevant for U-folds. A simple example of a U-fold in this formulation would simply be a torus bundle over the external manifold $M$ --- in the $\Gsix$ case, we would have a 5-dimensional external manifold that is the base for a $T^{27}$-bundle patched by $\Gsix$ transformations. For genuine U-folds, $q^M$ will change from patch to patch.

{\bf Acknowledgements:} ASA would like to thank Chris Hull for helpful discussions and for reading this manuscript and was supported by the EPSRC programme grant ``New Geometric Structures from String Theory'' (EP/K034456/1). 
CB thanks Emanuel Malek and Henning Samtleben for useful discussions.
CB is supported by an FWO-Vlaanderen postdoctoral fellowship, by the Belgian Federal Science Policy Office through the Interuniversity Attraction Pole P7/37 ``Fundamental Interactions'', by the FWO-Vlaanderen through the project G.0207.14N, by the Vrije Universiteit Brussel through the Strategic Research Program ``High-Energy Physics''.

\bibliography{../NewBib}
\end{document}